\documentclass{emulateapj}

\usepackage{natbib}
\shorttitle{ELVIS Observations of J1838--3427}
\shortauthors{Boyce et al.}

\begin{document}

\title{The Extragalactic Lens VLBI Imaging Survey (ELVIS).\\
I.\ A Search for the Central Image in the Gravitational Lens PMN~J1838--3427
}

\author{
Edward R.\ Boyce\altaffilmark{1,2},
Joshua N.\ Winn\altaffilmark{1},
Jacqueline N.\ Hewitt\altaffilmark{1}, and
Steven T.\ Myers\altaffilmark{2}
}

\email{eboyce@mit.edu, jwinn@space.mit.edu,\\ jhewitt@space.mit.edu, smyers@nrao.edu}

\altaffiltext{1}{Department of Physics and Kavli Institute for
Astrophysics and Space Research, Massachusetts Institute of
Technology, 77 Massachusetts Avenue, Cambridge MA 02139}

\altaffiltext{2}{National Radio Astronomy Observatory, PO Box O,
Socorro NM 87801}

\begin{abstract}
The Extragalactic Lens VLBI Imaging Survey (ELVIS) searches for
central images of lensed radio quasars, in order to measure the
central density profiles of distant galaxies.  Here we present
sensitive multi-epoch Very Long Baseline Array (VLBA) observations of
PMN~J1838--3427 at 8~GHz, with a $1~\sigma$ noise level of
$38~\mu$Jy~beam$^{-1}$. Based on the absence of a central image of the
background source at this level, we explore the possibilities for the
central matter distribution in the lens galaxy.  A power-law density
profile, $\rho \propto r^{-\gamma}$, must have $\gamma > 1.93$. Thus
the density profile is close to an isothermal profile ($\gamma=2$) or
steeper. The upper limit on any constant-density core in an otherwise
isothermal profile is $\lesssim5$~parsecs.  We also derive the constraints
on models in which the density profile is isothermal on kiloparsec
scales, but is allowed to have a different power law in the central
$\sim100$~parsecs. If the lens galaxy harbors a supermassive black
hole, the galaxy profile is allowed to be shallower, but for the
expected black hole mass the galaxy profile must still be close to
isothermal or steeper.
\end{abstract}


\section{Introduction} \label{intro}

The central regions of galaxies ($r<1$~kpc) are a topic of great
interest, but one for which it is difficult to compare the results of
observations and theoretical simulations. The observational difficulty
is especially severe for galaxies at cosmological distances. The
Extragalactic Lens VLBI Imaging Survey (ELVIS) is motivated mainly by
the desire to measure (or place interesting constraints on) the
central matter distribution in galaxies at significant redshift, by
searching for central images of gravitational lenses.  As this is the
first ELVIS publication, we elaborate on the theoretical and
observational situation.

On the theoretical side, cosmological dark matter simulations produce
dark matter halos with a universal density profile that goes as
$\rho\propto r^{-3}$ at large radii and $\rho\propto r^{-\gamma},
\gamma=1-1.5$ at small radii \citep{nav97a,moo99a}. The transition
occurs at a scale radius $r_s$, typically $10-30$~kpc for halos of
galactic mass. The main difficulty with interpreting the results for
galaxy interiors is that on scales $\sim10$~kpc and smaller, the
baryonic component is expected to modify the dark matter halo
significantly. Adiabatic contraction \citep{blu86a} and similar models
\citep{sel05a} model the contraction of dark matter under the
gravitation of a baryonic disk or bulge. The total matter profile
steepens and becomes close to isothermal ($\rho\propto r^{-2}$) on
scales of a few kpc \citep{kaz05a}. Similar results are found from
hydrodynamic simulations that use cooling, gas dynamics, star
formation and gravitation to model the dark matter and baryons within
individual halos. The baryonic matter dissipates energy and collects
at the halo center, becoming the dominant component inside radii
$1.5-5$~kpc \citep{gne04a,mac06a}.  The dark matter contracts inwards
under the gravitational influence of the baryons, again giving a total
matter profile that is close to isothermal. At present the smallest
scales probed by the hydrodynamic simulations are 0.3--1~kpc.

On the observational side, the density distributions of massive
galaxies can be directly probed through dynamical studies, at least in
the nearby universe.  \citet{ger01a} have modelled the dynamics of a
sample of large early-type galaxies, using photometry and kinematic
line profiles.  The dark matter fraction is only $10-40\%$ within the
effective radius $R_e$, while the rotation curves are flat on scales
larger than $0.2R_e$, indicating an isothermal density profile on
these scales.  Typically $R_e=4-10$~kpc \citep{kro00a}, so these
results agree with the simulations described above. The density
profile of an early type galaxy is isothermal at radii of a few kpc,
and baryons represent most of the mass inside this radius.

Projected surface brightness profiles of nearby galaxies
($z=0.002-0.005$) have been observed with the Hubble Space Telescope
\citep{lau95a}.  The angular resolution is 0\farcs1, which corresponds
to a physical size of $\sim10$~pc for the typical galaxy in this
sample.  The surface brightness profiles are well fit by a broken
power law, with steep outer exponents, shallower inner exponents, and
break radii between 10 and 1000~pc. Based on their inner profiles
$I(R)\propto R^{-\beta}$ the galaxies can be classified into two
populations: steep cusps with $\beta\sim1$ and flatter cores with
$\beta=0-0.3$. A surface density power law $\Sigma(R)\propto
R^{-\beta}$ corresponds to a density power law $\rho(r)\propto
r^{-\gamma}$, where $\gamma=\beta+1$.  Thus the luminous matter in a
cuspy galaxy has an isothermal distribution ($\Sigma(R)\propto
R^{-1})$ to within $\sim10$~pc of the galaxy's center, while the
luminous matter in a core galaxy breaks to a shallower profile at some
radius $\lesssim1$~kpc.  This is a good approximation to the total
mass profile, as the stars seem to represent most of the mass at these
radii.

Gravitational lenses provide information on the mass profiles of more
distant (and therefore younger) galaxies. For a lens galaxy at
$z=0.3-1.0$, the relative positions and fluxes of the bright lensed
images of a background source constrain the matter profile interior to
a few kpc from the lens galaxy center.  Detailed studies of $\sim20$
gravitational lenses find that the early type lens galaxies have
density profiles which are very close to isothermal on these scales
\citep{rus03a, koo06a}.  Distant early-type galaxies have similar
profiles to those nearby: they are isothermal at galactic radii of a
few kpc.

But what about the central few hundred parsecs?  Here, too,
gravitational lenses can help, through the properties of the ``central
image.''  In theory, a non-singular galactic profile produces an odd
number of images \citep{dye80a, bur81a}.  One image forms near the
center of the lens galaxy, where it is expected to be highly
demagnified by the large surface density at that position
\citep{nar86a}. Due to the demagnification, the faint central image is
rarely observed, leaving two or four bright images.  In cases where
the density profile is singular, with a central cusp that is stronger
than isothermal ($\rho \sim r^{-\gamma}$ with $\gamma \geq 2$), no
central image is produced even in theory.

While they are hard to observe, central images probe the inner
10--100~pc of very distant galaxies.  \citet{win03a, win04a} have
confirmed the existence of a central image produced by an isolated
lens galaxy in the lens PMN~J1632-0033, using radio observations.  We
discuss this object and its matter profile further in
Section~\ref{conclusions}.  \citet{ina05a} found a central image
generated by the combined profile of a cluster and a massive galaxy,
using the Sloan Digital Sky Survey and the Hubble Space Telescope (HST). 
The mass modelling here is dependent on both the galaxy and the cluster
profile.

Most galaxies host a super-massive black hole (SMBH) whose mass
correlates closely with the properties of the central stellar bulge
\citep{kor95a,mag98a}, particularly the velocity dispersion (the
$M-\sigma$ relation; \citet{fer00a,geb00a,tre02a}).  A SMBH in the
lens galaxy affects the central image: it can destroy the central
image, or split the central image into two images, one of which is
directly attributable to the black hole \citep{mao01a,bow04a}.  In the
latter case, the properties of the central-image pair could allow for
the measurement of the black hole mass in an ordinary galaxy at
significant redshift \citep{rus05a}.

Based on the surface-brightness profiles of nearby early-type galaxies
measured with HST, \citet{kee03a} predicted the distribution of core
image magnifications. He found a broad distribution, from $10^{-4.5}$
to $10^{-1}$, with a most probable magnification of approximately
$10^{-2.5}$.  Adding a SMBH had little effect on central-image
detectability: the magnification of the central image was strongly
affected only when the magnification of the galaxy alone was already
very low.  \citet{rus01a} placed limits on central mass distributions
using radio observations of six lenses from the Cosmic Lens All-Sky
Survey (CLASS; \citet{bro03a, mye03a}).  
Their basic result was that if the density profiles in the
lens galaxies are taken to be power laws in radius, then they must be
nearly isothermal or steeper.  The constraints were somewhat weaker
when a SMBH was included.

This paper is organized as follows. Section~\ref{elvis} describes the
design of our survey. Section~\ref{j1838} presents new observations of
our first target, PMN~J1838--3427. Section~\ref{models} presents
models for the density profile in the lens galaxy of this system, and
the final section summarizes the results.

\section{ELVIS} \label{elvis}

The Extragalactic Lens VLBI Imaging Survey (ELVIS) will involve
sensitive, high-angular-resolution radio observations of many of the
known cases of gravitational lensing of a radio-loud quasar. The
traditional reasons to conduct such observations are to confirm cases
of gravitational lensing, and to observe correspondences between
lensed radio jets in order to refine models of the lens galaxy. 
ELVIS is the first survey (to our knowledge) motivated by the search
for central images. As such, our highest-priority targets are those
that are most favorable for central-image hunting: radio-loud,
asymmetric two-image lenses.

The asymmetric two-image lenses (those with a large magnification
ratio between the two bright quasar images) are best because for those
systems, the mean magnification of the central image is generally the
largest, for a given lens galaxy \citep{mao01a,bow04a}. In a highly
asymmetric system, the angular separation between the lens galaxy and
the unlensed source is nearly as large as possible, while still being
close enough to produce multiple images. This results in a central
image that is as distant as possible from the lens galaxy center. More
symmetric systems (especially those that produce four bright images)
have central images located closer to the galaxy center, where the
surface density is larger and the degree of demagnification is
consequently greater.

Observing at radio wavelengths is desirable to avoid extinction by
dust within the interstellar medium of the lens galaxy. Attentuation
by plasma effects is also possible, but can be minimized by observing
at a high enough radio frequency (typically $\ge 5$~GHz). Moreover, an
optical image could easily be lost in the starlight of the lens
galaxy, while even a demagnified radio image will be brighter than a
typical radio-quiet lens galaxy. The angular separation between the
central image and the other images, or the lens galaxy center, is
likely to be 100~mas or less, so at the widely used radio frequencies
of 1-10~GHz, very-long-baseline interferometry (VLBI) is needed.

\begin{table*}
\begin{center}
\caption{Details of the VLBA observations.
\label{vlba}}
\begin{tabular}{crrrrrr}
\tableline\tableline
Date & Image A & flux density & Image B & R.A.$_{\rm A}$-R.A.$_{\rm B}$ & 
dec.$_{\rm A}$-dec.$_{\rm B}$ & Blank field\\
& point source & total & flux density & & & rms\\
& (mJy) & (mJy) & (mJy) & & & ($\mu$Jy~beam$^{-1}$)\\
\tableline
2000 Oct 07 & 206.8 & 219.2 & 9.3 & 0\farcs09747 & 0\farcs99130 & 86\\
2000 Oct 30 & 204.3 & 213.4 & 16.2 & 0\farcs09747 & 0\farcs99125 & 88\\
2000 Dec 17 & 222.4 & 232.3 & 12.8 & 0\farcs09746 & 0\farcs99122 & 95\\
2000 Dec 18 & 202.7 & 214.4 & 12.3 & 0\farcs09747 & 0\farcs99121 & 98\\
2000 Dec 22 & 209.7 & 209.7 & 13.4 & 0\farcs09745 & 0\farcs99118 & 93\\
2001 Mar 24 & 207.5 & 219.5 & 18.3 & 0\farcs09744 & 0\farcs99111 & 79\\
\tableline
\end{tabular}
\tablecomments{Details of the individual observing epochs.
The beam size varied slightly between epochs, with average values
of $3.5\times1.3$~mas, and the position angle was always within $1\degr$ of
zero. All epochs excluded the Brewster antenna due to low elevations.
The December epochs lacked the North Liberty VLBA antenna, 
December 17 and December 18 substituted a single Very Large Array (VLA) 
antenna for the Pie Town VLBA antenna.}
\end{center}
\end{table*}

The search for central images is a task that is well-matched to recent
advances in VLBI technologies. New high-bandwidth recorders and
digital back-ends have increased typical data rates by a factor of 4
in recent years, with even greater improvements expected in the next
few years. The effect is a considerable sensitivity boost for those
experiments that can take advantage of the increased bandwidth. The
European VLBI Network now routinely records at 1~Gb~s$^{-1}$. The
U.S.\ National Radio Astronomy Observatory does not yet generally
deploy recorders with bandwidths this wide, but it recently began
devoting approximately 300 hours per year to simultaneous observing
with the 10-station Very Long Baseline Array (VLBA) and several other
large antennas such as the phased Very Large Array (VLA), Arecibo, the
Green Bank Telescope, and the Effelsberg telescope. The thermal noise
level with this ``High Sensitivity Array'' (HSA) is often 10-20 times
lower than a traditional VLBA observation of similar
duration. Gravitational lens central image searches are perfect
projects for sensitive observations with heterogeneous arrays. The
search for a faint central image in an otherwise blank region is a
problem limited mainly by the thermal noise level, rather than the
need for complete coverage of the Fourier plane. The typical radio
lens field has two or four bright, compact sources (the bright images
of the lensed quasar), and no extended structure. The bright images
provide in-beam phase calibration sources.  The simplicity of the
source structure (multiple, isolated point sources) makes it easier to
calibrate and image the data from the heterogeneous array.

For all these reasons, the first ELVIS targets are radio-loud quasars
lensed by a single galaxy to produce two bright images with a flux
ratio exceeding 5:1.  Two lenses that fit this description,
PMN~J1632-0033 \citep{win02a} and CLASS~B1030+074 \citep{xan98a}, have
already been the subjects of sensitive VLBI observations by other
groups. In the former case, good evidence for a central image was
found \citep{win04a}.  We are in the process of obtaining data from
long-duration VLBA observations, or shorter-duration HSA observations,
of the other known radio lenses meeting these criteria, many of which
are taken from CLASS.

Here we present our first target, PMN~J1838--3427.  This gravitational
lens was discovered by \citet{win00a}, as part of a survey for
radio-loud gravitational lenses in the southern sky. It has a two
bright images separated by 1\farcs0 and with a flux ratio of 14:1. The
redshift of the source quasar is at $z_{\rm S}=2.78$. A spectroscopic
redshift for the lens galaxy has not been measured, despite several
attempts. It is a difficult measurement mainly because the optical
light of the lens galaxy is blended with that of the fainter lens
image, and also because the system resides in a crowded field at low
galactic latitude. \citet{win00a} estimated $z_{\rm L}=0.36\pm0.08$,
the range of redshifts for which the lens galaxy photometry is
consistent with the fundamental plane relation. The source was
monitored at 9~GHz with the Australia Telescope Compact Array
\citep{win04b}, as discussed in Section~\ref{scintil}.

\section{Observations of J1838-3427} \label{j1838}

We observed PMN~J1838-3427 (hereafter, J1838) with the ten antennas of
the NRAO Very Long Baseline Array (VLBA), on 6 different epochs between
2000 October and 2001 March. At each epoch, the duration of the
observation was 5 hours. We observed right circular polarization with
a central frequency of 8.415~GHz. We used 2 bit sampling at 16
Msamples~s$^{-1}$ for each of 8 channels, giving a total data rate of
256 Mb~s$^{-1}$. Data from antennas that were observing at an
elevation of less than 10 degrees were excluded (this included all of
the data from the Brewster antenna). In addition, the North Liberty
antenna was not in use for two epochs, and was excluded from a third
epoch for which its data were very noisy and degraded the image. Thus,
3 epochs used 9 antennas and 3 epochs used 8 antennas. In each
observation, the time on J1838 was 4.5 hours and the estimated thermal
noise limit (based on the collecting area and receiver
characteristics) was $\sim70~\mu$Jy~beam$^{-1}$.

For each epoch the data were amplitude calibrated in AIPS, following
standard procedures. Because image A of the gravitational lens is
fairly bright, with a flux density of $\sim200$~mJy, we did not need
to perform phase referencing.  The initial phase solution was derived
from a fringe fit to a point-source model centered at the location of
image A, using a solution interval of 2 minutes.  We then reduced the
data with nine self-calibration cycles, each cycle consisting of
imaging, phase-only self-calibration with a 0.5-1.5 minute solution
interval, imaging, and phase and amplitude self-calibration with a 15
minute solution interval. For each new cycle, smaller CLEAN components
were included.

\begin{figure}
\includegraphics[scale=0.66]{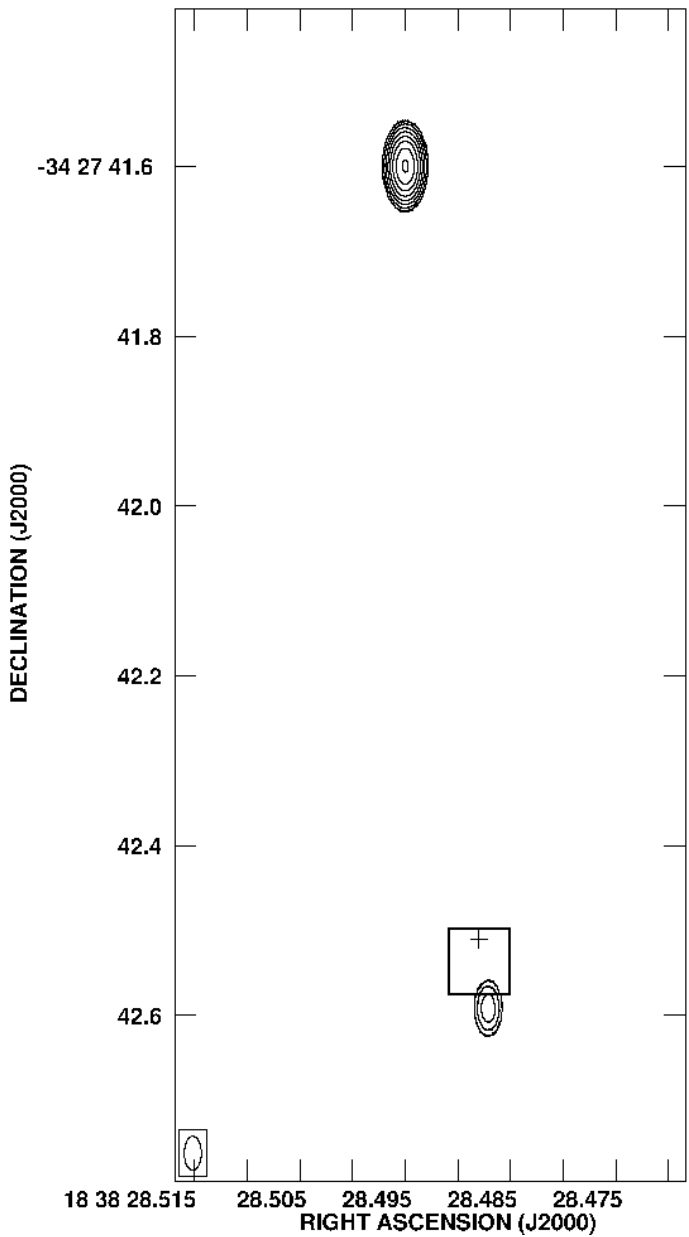}
\caption{ Radio map of the two images of gravitational lens
PMN~J1838--3427, made with the VLBA at 8~GHz.  Data are from the first
epoch. The restoring beam ($40\times 20$~mas) was chosen to be much
larger than the naturally weighted beam in order to show both images
on a single map. Contours begin at 1.5~mJy~beam$^{-1}$ and increase by
factors of 2. The J2000 radio positions were not determined from our
observations; they were assumed from earlier VLA imaging. The cross
marks the location of the lens galaxy detected using HST/WFPC2; its
position is based on the offset from the bright quasar images in the
WFPC2 images. The box shows the central image search region.
\label{wholefield}}
\end{figure}

\begin{figure}
\plottwo{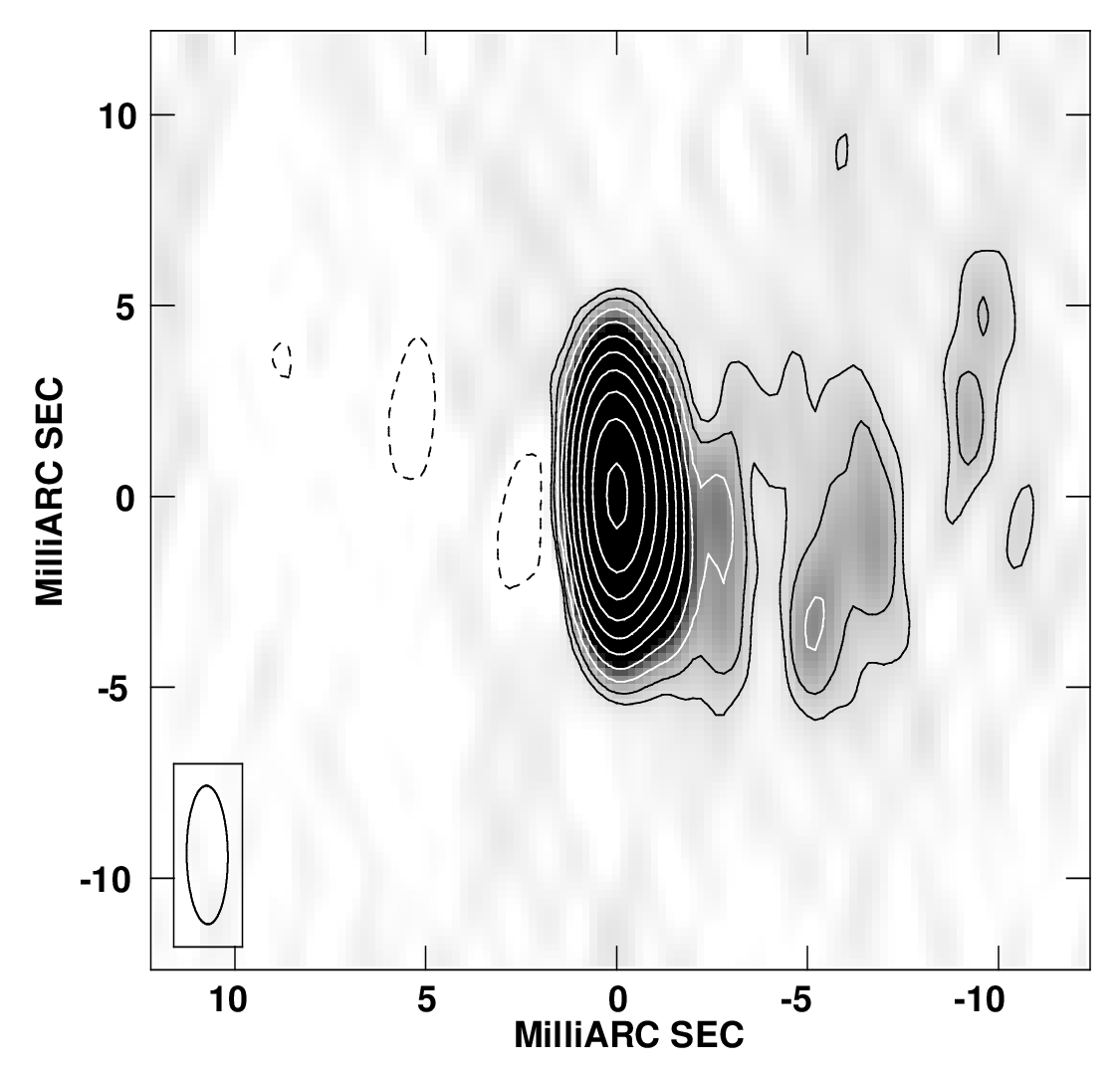}{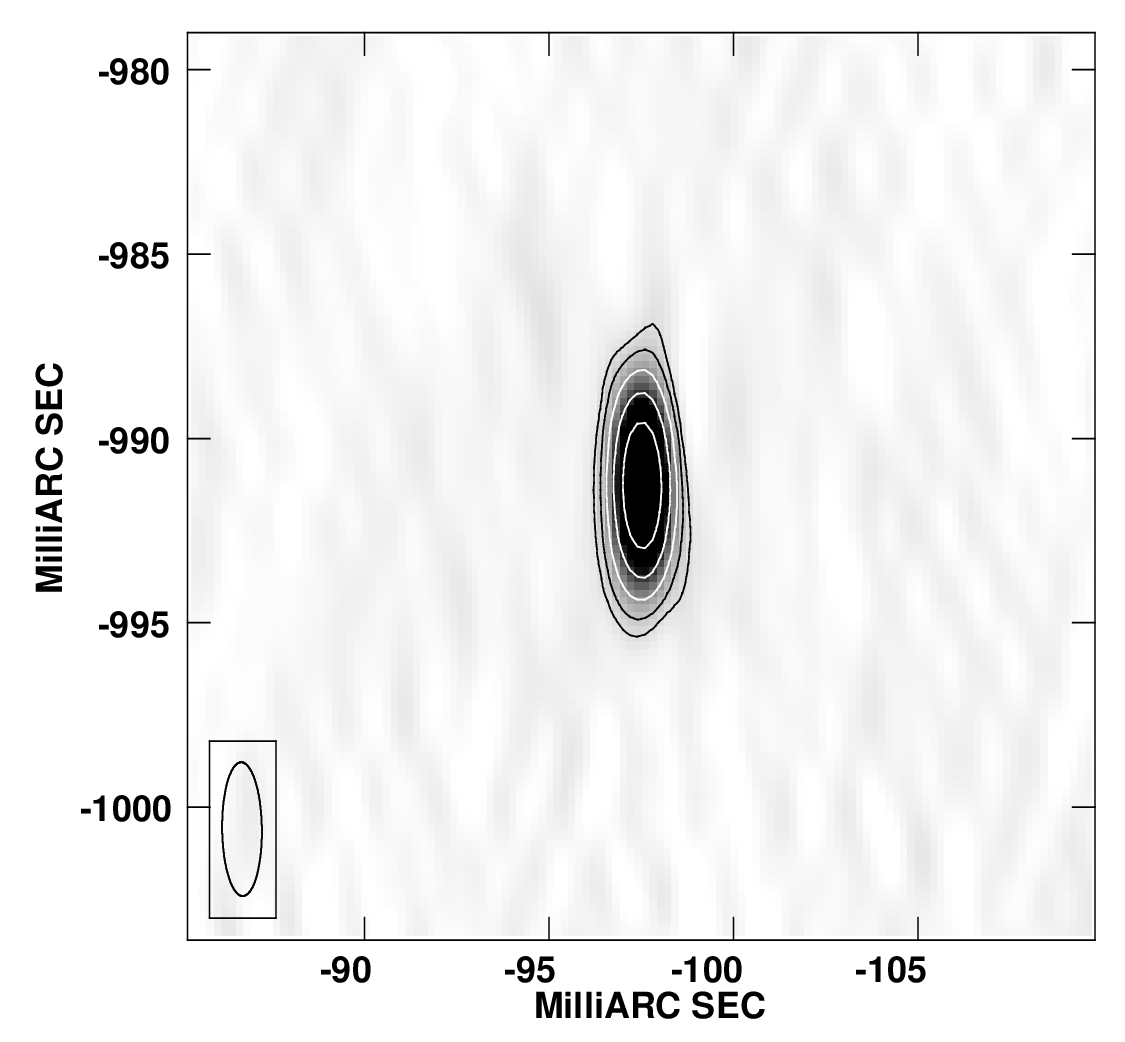}
\caption{ (left) Radio map of image A.  (right) Radio map of image B.
Both maps were made from the first epoch of VLBA imaging at 8~GHz.
Coordinate offsets are from the phase center at (J2000) $18^{\rm
h}34^{\rm m}28\fs495, -34\degr27\arcmin41\farcs60$.  Contours begin at
$300~\mu$Jy~beam$^{-1}$ and increase by factors of 2 (the blank field
rms near image B was $86~\mu$Jy~beam$^{-1}$).  The synthesized beam of
$3.6\times1.1$~mas is shown at the lower left.  Note the extended
emission to the west of the main point source in image A.
\label{ABims}}
\end{figure}

The brightest image A appeared as a point source of flux
$\sim210$~mJy, with $\sim10$~mJy of extended emission to the west,
while image B appeared as a point source, varying between
9~and~18~mJy.  Maps from the first epoch are shown in
Figure~\ref{ABims}, while details of each observing run are presented
in Table~\ref{vlba}. The single epoch 5~GHz VLBA map of \citet{win00a}
shows two bright point sources at the same positions, with diffuse
emission to the west of image A. The fraction of the total image A
flux density in this extended component is $\sim10$\% at 5~GHz and
$\sim5$\% at 8.4~GHz, so the extended emission has a steeper spectral
index than the point source.
 
The flux densities of the two bright images varied between the epochs,
and the ratio of these flux densities also varied.  The image A to
image B flux density ratio in our maps varies from 12 to 24, with an
average of 15. This ratio averaged 14.6 in VLA observations and 10.6
in the the previous 5~GHz VLBA observations \citep{win00a}.  The lens
images appear to undergo large variations in intensity.  The two most
obvious possible explanations are intrinsic source variation, and
interstellar scintillation (see Section~\ref{scintil}).

\begin{figure}
\plotone{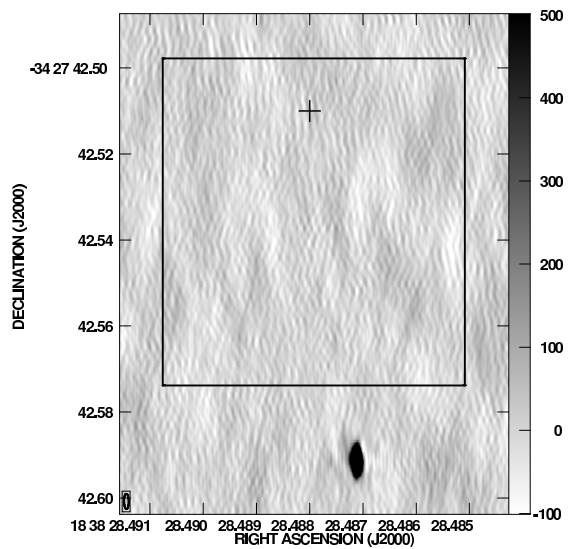}
\caption{ Radio map of the central image search region, from the
combined VLBA 8~GHz map of all six epochs. The wedge at right shows
the grey scale in $\mu$Jy~beam$^{-1}$.  The J2000 radio positions were
not determined from phase referencing; they were assumed from earlier
VLA imaging.  The cross marks the location of the lens galaxy detected
using HST/WFPC2. Its position is based on the offset from the bright
quasar images in the WFPC2 images.  Image B is obvious near the bottom
of the map.  The central image would form at the light travel time 
maximum located between the lens galaxy and the bright image B, 
so we searched for the central image in the boxed area.
\label{censearch}}
\end{figure}

\begin{figure}
\plotone{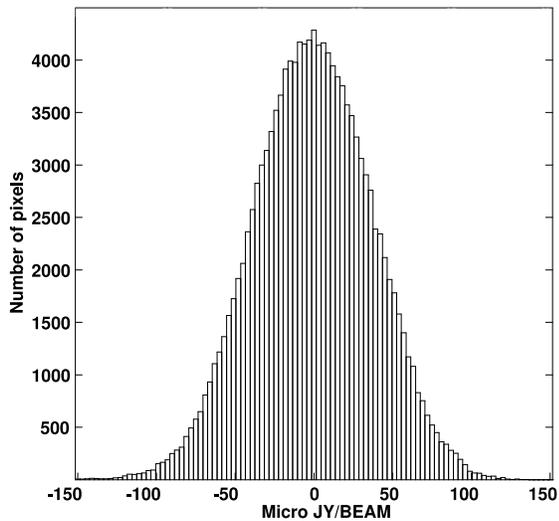}
\caption{ Histogram of surface brightness values for all pixels within the
central-image search region. The pixel size is 0.2~mas, which is
smaller than the typical beam size of $3.5\times1.3$~mas. The
distribution of surface brightness is approximately Gaussian, with a mean of
$-0.4~\mu$Jy~beam$^{-1}$ and a standard deviation of
$38~\mu$Jy~beam$^{-1}$.
\label{cenhist}}
\end{figure}

No additional sources of radiation were seen in any epoch. We examined 
a large region between the location of the lens galaxy and the bright 
image B, where a central image would be expected.
\footnote{Gravitational lens images form at extrema of the time delay 
surface, which is a combination of the geometric delay and the lens 
galaxy potential determined by its density profile. The central image 
forms near the lens galaxy potential maximum. The geometric delay 
shifts the overall maximum towards the saddle point image B, on the 
opposite side of the lens galaxy from the source. The central image 
will thus form between the lens galaxy and image B, the fainter of 
the two bright images.}
We extended this region slightly to the other side of the lens galaxy 
to allow for the uncertainty in the optical position.
From epoch to epoch the noise varied from
$79~\mu$Jy~beam$^{-1}$ to $98~\mu$Jy~beam$^{-1}$ (Table~\ref{vlba}),
which is within 20-30\% of the expected thermal noise limit. The small
but significant excess over the thermal noise limit is expected, given
the southerly declination of the target. When the six maps were
co-added, the central-image search region remained blank (see
Figure~\ref{censearch}) and the noise decreased by
$\sim$$\sqrt{6}$. The distribution of surface brightness in the
pixels of the final map is roughly Gaussian, with a mean of
$-0.4~\mu$Jy~beam$^{-1}$ and a standard deviation of 
$38~\mu$Jy~beam$^{-1}$ (see Figure~\ref{cenhist}).

To quote an upper limit, we considered directly the surface 
brightness distribution and took the 99th percentile to be the 99\% 
limit on the central image flux density. This gave an upper limit on
the central image flux density of $S_{\rm C}<83~\mu$Jy, and a lower 
limit on the magnification ratio of the brightest image to the central 
image of $S_{\rm A}/S_{\rm C}>2500$, both with 99\% confidence. For 
comparison, the 5~GHz VLBA maps of \citet{win00a} reached an rms
of $190~\mu$Jy~beam$^{-1}$, and measured a flux density of 
$S_A=145$~mJy for the brightest image. Assuming a Gaussian 
distribution for the blank field errors, the old 99\% limits were 
$S_C<470~\mu$Jy and $S_{\rm A}/S_{\rm C}>310.$ 

\subsection{Scintillation} \label{scintil}

We found the flux densities $S_{\rm A}$ and $S_{\rm B}$ of the two
bright quasar images to vary from epoch to epoch.  The ratio $S_{\rm
A}/S_{\rm B}$ also varied significantly, demonstrating that the
variations are not due only to inconsistencies in the flux density
scale.  Image B showed a higher fractional variation than image A
(Table~\ref{vlba}). This is similar to what was observed by
\citet{win04b}, who monitored this object for 4 months with the
Australia Telescope Compact Array (ATCA) at 9~GHz.  Over the course of
the campaign, the fractional variation in the flux density of image A
was 4\%, and that of image B was 8\%.

The lens is located at a low galactic latitude ($b_{\rm II}=-12\fdg5)$, 
and may undergo scintillation due to the Milky Way's
interstellar medium. The root-mean-squared amplitude due to
scintillation is inversely proportional to the angular size of the
source \citep{wal98a, wal01a}, while the angular size of each lens
image is proportional to $S^{1/2}$, since gravitational lensing
conserves surface brightness. Thus the rms amplitude of scintillation
variations is proportional to $S^{-1/2}$, and the fainter image B
would be more affected by interstellar scintillation than image A.  In
contrast, intrinsic variability should produce the same fractional
variation in each component (though the variations would appear with
time lags due to the geometric and Shapiro delays).  The greater
fractional variation of image B that was observed with ATCA, and in
our own observations, supports the scintillation hypothesis.

The central image, if it exists, has a much smaller flux density and
angular size than either of the bright images.  Taking $S_{\rm
A}/S_{\rm C} > 2500$, scintillation would cause fractional variations
in $S_{\rm C}$ of order unity. No central radio source was seen in any
individual observation. Scintillation, if present, did not magnify the
central image above the detection limit for any single epoch
($\sim200~\mu$Jy). In what follows, we use the blank field of the
combined map to define our upper limit on the central image flux
density, but we note that scintillation could be a major source of
systematic error in this determination.

\section{Mass Models} \label{models}

\subsection{Cuspy Matter Profiles} \label{cuspy}

With this new and more stringent upper limit on the flux density of
the central image, we can restrict the possibilities for the central
density profile of the J1838 lens galaxy. A simple and realistic model
for the mass distribution of a massive galaxy is a broken power law
(see Section~\ref{intro}). We adopt the broken power law density
profile of \citet{mun01a}, in which the surface density and the
deflection angle are given by analytic expressions. This profile
varies as $\rho\propto r^{-n}$ at large radii, and as $\rho\propto
r^{-\gamma}$ at small radii, with a break at radius $r_b$:
\begin{equation}
\rho(r) = \frac{\rho_0}{r^\gamma} 
\frac{1}{(1+r^2/r_b^2)^{(n-\gamma)/2}}.
\end{equation} 

Since galactic profiles are approximately isothermal on scales of a
few kpc (see Section~\ref{intro}) and we have few constraints, we
apply the method used by \citet{win03a} in the analysis of
PMN~J1632--0033 and fix the outer power law to be isothermal ($n=2$).
We then explore the constraints on the break radius $r_b$ and inner
power law index $\gamma$.  In the limit of a pure isothermal sphere,
the central image vanishes; thus, as $r_b$ goes to zero or as $\gamma$
goes to 2, the central image flux density approaches zero.  We use a
spherical galaxy model, and account for non-sphericity in the
profile with an external shear at an arbitrary position angle. The
10 free parameters of our model are the positions of the lens galaxy
and the source, ($x_{\rm G}, y_{\rm G}, x_{\rm S}, y_{\rm S}$), the
source flux ($S_{\rm S}$), the mass parameter 
($b=(2\pi r_b\rho_0)/(\Sigma_{cr})$)
\footnote{The quantity $\Sigma_{cr}=(c^2/4\pi G)(D_S/(D_LD_{LS}))$ is
the critical surface density for lensing. The quantities $D_L$, $D_S$
and $D_{LS}$ are the angular diameter distances from the observer to
the lens, from the observer to the source and from the lens to the
source, respectively.} the shear and its position angle ($s,
\theta_s$), the break radius ($r_b$) and the inner power law index
($\gamma$).

The data provide 9 observables: the two sky coordinates for each of
the lens galaxy, bright image A and bright image B; the flux densities
for image A and image B ($S_{\rm A}, S_{\rm B}$); and an upper limit
on the flux density of the central image C ($S_{\rm C}$).  The galaxy
coordinates and their uncertainties were taken from the Hubble Space
Telescope Wide Field Planetary Camera 2 (WFPC2) photometry by
\citet{win00a}.  The radio coordinates and flux densities for images A
and B were measured in each of the six VLBA maps and the mean values
of these observables were taken as model constraints. We considered
only the bright peak of image A, and ignored the diffuse emission to
the west, as we consider only the point source magnifications.  Since
both the positions and fluxes of the bright images may be perturbed by
substructure in the lens galaxy, we adopted errors of 3~mas in the
positions and 20\% in the flux densities (cf.\
\citet{mao98a,kee98a,koc04a}). The constraint on $S_{\rm C}$ was taken
from the rms noise in a large region between the lens galaxy position 
and image B in the combined VLBA image.  The distribution of surface 
brightness in the pixels of this blank region was approximately 
Gaussian, with a mean of $-0.4~\mu$Jy~beam$^{-1}$ and a
standard deviation of $38~\mu$Jy~beam$^{-1}$ (see Figure~\ref{cenhist}).  
We took this to be a null detection of the central image with a 
$1~\sigma$ error of $38~\mu$Jy~beam, and set
$S_{\rm C} = 0\pm38~\mu$Jy for the modelling.  These constraints are 
summarized in Table~\ref{modcons}.  All of the observables were
assumed to obey Gaussian statistics, and $1~\sigma$ error bars
are quoted.

\begin{table}
\begin{center}
\caption{Lens model constraints.\label{modcons}}
\begin{tabular}{cr}
\tableline\tableline
Observable & Value\\
\tableline
R.A.$_{\rm A}$ & $0\farcs000\pm0\farcs003$\\
decl.$_{\rm A}$ & $0\farcs000\pm0\farcs003$\\
R.A.$_{\rm B}$ & $-0\farcs9912\pm0\farcs003$\\
decl.$_{\rm B}$ & $-0\farcs0975\pm0\farcs003$\\
R.A.$_{\rm G}$ & $-0\farcs911\pm0\farcs006$\\
decl.$_{\rm G}$ & $-0\farcs085\pm0\farcs006$\\
$S_{\rm A}$ & $208.3\pm41.8$ mJy\\
$S_{\rm B}$ & $13.7\pm2.7$ mJy\\
$S_{\rm C}$ & $0.000\pm0.038$ mJy\\
\tableline
\end{tabular}
\tablecomments{Constraints on the lens models, with $1~\sigma$ errors.
All positions are offsets from image A.
Galaxy postions are from the WFPC2 photometry in \citet{win00a}, other
constraints are from the VLBA data.
}
\end{center}
\end{table}

\begin{figure}
\plotone{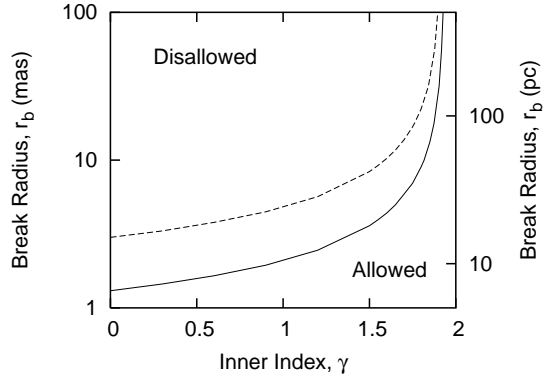}
\caption{Models for the lens galaxy of PMN~J1838--3427, using the new
data presented in this paper (solid line) and the previous
data from \citet{win00a} (dashed line). The matter profile is isothermal 
($\rho\propto r^{-2}$) at large radius, and goes as 
$\rho\propto r^{-\gamma}$ at small radius, with the break at radius 
$r_b$. The inner power law index and break radius are allowed to vary. The
lines mark the $\chi^2=5.99$ boundary between
the allowed region (below and to the right) and the disallowed region
(above and to the left). The break radius in parsecs is calculated for
$z_{\rm L}=0.36$.
\label{old_new}}
\end{figure}

We model the lens with the {\it gravlens} package by
\citet{kee01a}. For the conversion from angular units to physical
units, we assume $z_{\rm L}=0.36$ and a $\Lambda$CDM cosmology with
$H_0=70~{\rm km~s}^{-1}~{\rm Mpc}^{-1}, \Omega_m=0.3,
\Omega_\Lambda=0.7$.  With 10 free parameters and 9 measurements, one
would expect a one-dimensional locus of allowed points in parameter
space. In fact, because the observable $S_{\rm C}$ is an upper limit,
the data define a one-dimensional boundary between an allowed region
and a disallowed region in parameter space. In Figure~\ref{old_new},
we plot the projection of this plane in the $r_b,\gamma$ parameter
space that characterizes the inner density profile.  Values of the
other parameters are given in Table~\ref{isoparms} for the isothermal
case ($r_b=2$ or $\gamma=0$), and these values change by
$\lesssim10\%$ over the region of $r_b,\gamma$ parameter space we
investigated.

\begin{table}
\begin{center}
\caption{Lens model parameters for the isothermal case. \label{isoparms}}
\begin{tabular}{cr}
\tableline\tableline
Parameter & Value\\
\tableline
$b=\theta_{\rm E}$ & 0\farcs528\\
$x_{\rm G}$ & -0\farcs0850\\
$y_{\rm G}$ & -0\farcs9109\\
$x_{\rm S}$ & -0\farcs0187\\
$y_{\rm S}$ & -0\farcs4747\\
$S_{\rm S}$ & 80.3 mJy\\
$s$ & 0.0649\\
$\theta_s$ & $-18\fdg1$\\
\tableline
\end{tabular}
\tablecomments{The best fit values of the lens model parameters for the
isothermal case (when $r_b=0$ or $\gamma=2$).
}
\end{center}
\end{table}

For a sufficiently small break radius $r_b$ or sufficiently steep
inner index $\gamma$, the model fits the data perfectly.  As $r_b$
increases or $\gamma$ decreases, the fits become progressively worse
and $\chi^2$ becomes large.  Most of the disagreement is due to an
unacceptably large $S_{\rm C}$. Near the boundary of the allowed 
region the best fits give slight variations in the calculated 
values for the other observables, but these are generally less than 
the $1~\sigma$ errors. The 95\% confidence region ($\chi^2<5.99$ for 
two free parameters, cf.\ \citet{bev03a}) lies below and to the 
right of the solid line. The matter distribution is nearly 
isothermal: a flat core with $\gamma=0$ must be very small 
($r_b \lesssim 1.5{\rm mas} \approx 5$~pc), and a large core with 
$r_b = 100{\rm mas} \approx 500$~pc has a steep inner profile 
($\gamma \gtrsim 1.92$).

For comparison, Figure~\ref{old_new} includes the region of parameter 
space allowed by the less stringent observational constraints of 
\citet{win00a}. The new data reduce the maximum size of a flat core 
by a factor of $\sim2.5$, and increase the minimum inner index of a 
large core by $\sim0.05.$

\subsection{Cuspy Matter Profiles with Black Holes} \label{smbh}

Most elliptical galaxies host super-massive black holes at their
centers \citep{kor95a,mag98a}: this additional point mass steepens the
overall central profile and may demagnify the central image
\citep{mao01a}. A realistic galaxy model should include a central
black hole. As seen in the previous section, even an interesting
smooth profile is underconstrained.  To avoid adding extra degrees of
freedom to the model, we add a black hole with a fixed mass given by
the $M-\sigma$ relation \citep{fer00a,geb00a,tre02a}, and determine
the resulting effect.

For an early-type lens galaxy, most of the mass will be in the bulge
component and so the total mass responsible for the lensing is a good
estimate of the overall velocity dispersion.  To estimate $\sigma$, we
determine the best-fitting value of the Einstein radius using a model
consisting of an isothermal sphere and external shear, and then use
the relation $\theta_{\rm E}=4\pi(D_{\rm LS}/D_{\rm
S})(\sigma^2/c^2)$.  For the broken power law model, the mass
normalization parameter $b=(2\pi r_b\rho_0)/(\Sigma_{cr})$ is
equivalent to $\theta_{\rm E}$ and this parameter varies by $<10\%$
over the allowed region in the broken power law models, so the
isothermal $\theta_{\rm E}$ is a good estimate of the true velocity
dispersion even if the true density profile is not exactly isothermal.

Given this estimate for $\sigma$, we calculate the black hole mass 
$M_{\rm bh}$ using the relation from \citet{tre02a},
\begin{equation}
M_{\rm bh}=1.35\times10^8~M_\odot(\frac{\sigma}{200~{\rm km~s}^{-1}})^{4.02}.
\end{equation}
For J1838, $\theta_{\rm E}=0\farcs523$, $\sigma=154~{\rm km~s}^{-1}$
and $M_{\rm bh}=4.7\times10^7~M_\odot$.  The scatter in the $M-\sigma$
relation is 0.3~dex \citep{tre02a}, so a plausible range for the black
hole mass is $2.5\times10^7-1.0\times10^8~M_\odot$.
 
\begin{figure}
\plotone{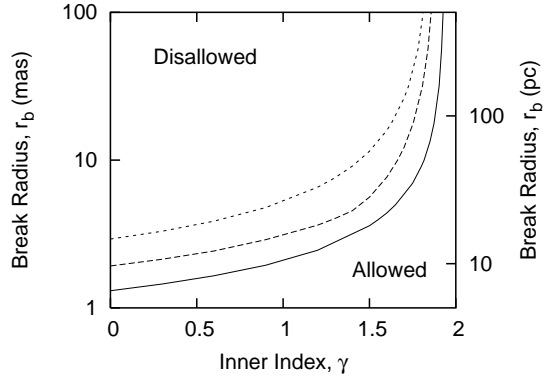}
\caption{Models for the lens galaxy of PMN~J1838--3427, using the new data
presented in this paper and including the effects of a super-massive black
hole.  The matter profile is isothermal ($\rho\propto r^{-2}$) at large 
radius, and goes as $\rho\propto r^{-\gamma}$ at small radius, with the 
break at radius $r_b$. To the smooth matter profile we add no black hole 
(solid line), a $5\times10^7~M_\odot$ black hole (long-dashed line), or a
$10^8~M_\odot$ black hole (short-dashed line).  These black hole
masses are the central and largest values expected, given the velocity
dispersion of the galaxy. The inner power law index and break radius
are allowed to vary. The lines mark the $\chi^2=5.99$ boundary between
the allowed region (below and to the right) and the disallowed region
(above and to the left). The break radius in parsecs is calculated for
$z_{\rm L}=0.36$. When no black hole is present, a central image
always forms within the parameter ranges plotted. In the allowed
region, the central image is below our detection limit.  When a black
hole is present, the central image is either eliminated or split into
a core image and SMBH image. In the allowed region, either there are
no central images or both central images have flux densities below our
detection limit.
\label{rb_gamma}}
\end{figure}

We rerun the calculations from Section~\ref{cuspy}, but this time
include a point mass fixed at the center of the lens galaxy.  We
consider a black hole with mass $5\times10^7~M_\odot$, in the center
of the expected range, and a black hole with the mass $10^8~M_\odot$,
at the top of the expected range.  The black hole either eliminates
the central image completely, leaving only the two bright images, or
creates two faint central images, the core and SMBH images, for a
total of four images.  The allowed regions, again defined by
$\chi^2<5.99$, are shown in Figure~\ref{rb_gamma}.  Below and to the
right of the allowed lines there are no central images, or both the
core and SMBH images are sufficiently faint that they would not have
been detected in our observations. The long-dashed line is for the
$5\times10^7~M_\odot$ black hole, the short-dashed line is for the
$10^8~M_\odot$ black hole.

The separation of the core and SMBH images is $2-5$~mas, so these might 
appear as a single unresolved component if aligned with the long axis 
of the beam. The core image flux density is greater than the SMBH 
image flux density by a factor of $15-100$, so the core image flux 
density makes a much larger contribution to $\chi^2$. For models 
near the boundary of the allowed region, combining both central images 
into a single unresolved component would raise $\chi^2$ by 
$< 0.6$, and the allowed region would be only slightly more
constrained.

The SMBH allows the galaxy profile to be shallower, to some
extent. For fairly flat cores ($\gamma\lesssim1.5$) the maximum break
radius increases by about 50\% for the $5\times10^7~M_\odot$ black
hole, and by a factor of $2.5-3$ for the $10^8~M_\odot$ black hole.  For
larger, cuspy cores, the minimum inner slope is reduced by $\sim0.1$
for the $5\times10^7~M_\odot$ black hole and by $\sim0.2$ for the
$10^8~M_\odot$ black hole.  While the central smooth profile remains
close to isothermal, it is allowed to be somewhat flatter. The cusp in
the profile from the black hole makes the central mass profile
sufficiently steep even if the smooth component is shallower.

\subsection{Comparison with Other Central-Image Modelling} \label{previous}

\citet{win03a} modelled PMN~J1632-0033, the lens with a confirmed
central image, using the same cuspy profile as this work. Given that
the central image is detected with a definite flux, the allowed region
in the break radius and inner index space becomes a narrow band (see
\citet{win03a}, Figure~6).  A flat core has a fixed small size, and a
larger core is forced to a particular value of the inner index.  The
allowed band for J1632 is similar to the boundary of the allowed
region for J1838: a central image of definite flux density selects a
narrow band giving that value for $S_{\rm C}$, while an upper limit on
the central flux density selects a half plane bounded by the band for
the maximum allowed $S_{\rm C}$.

Other central image modelling has focused on single power law models.
\citet{win03a} also considered a single power law model $\rho\propto
r^{-\gamma}$ for the density profile of J1632, finding
$\gamma=1.91\pm0.02$.  This modelling did not include a central black
hole.  \citet{rus01a} modelled the gravitational lenses CLASS
B0739+366 and B1030+074 using a single power law model for the
projected surface density $\Sigma(R)\propto R^{-\beta}$, both with and
without a central black hole with a mass inferred from the $M-\sigma$
relation. Their only constraint was the limit on the bright image to
central image magnification ratio, the observable which dominates
$\chi^2$ in our modelling.  Their limits on the smooth profile were
$\beta>0.85$ for B0739+366 and $\beta>0.91$ for B1030+074, while the
respective limits were $\beta>0.84$ and $\beta>0.83$ when a black hole
was included.

Modelling J1838 as a single power law in the density $\rho\propto
r^{-\gamma}$, we find $\gamma>1.93$ for the smooth profile,
$\gamma>1.86$ when we include a central black hole of
$5\times10^7~M_\odot$ and $\gamma>1.82$ when we include a central
black hole of $10^8~M_\odot$ (the central and largest values expected
from the velocity dispersion).  Again, we note that a surface density
index $\beta$ is equivalent to a density index $\gamma=\beta+1$, so
these 4 lenses are similar. The profile is slightly shallower than
isothermal for J1632--0033, and the profiles are slightly shallower
than or steeper than isothermal for J1838--3427, B0739+366 and
B1030+074.

\section{Conclusions} \label{conclusions}

We have presented new VLBA data for the gravitational lens
PMN~J1838--3427, and set a stringent upper limit of $83~\mu$Jy (99\%
confidence) on the observed flux density of the central image. The
corresponding limit on the brightest image to central image
magnification ratio is $S_{\rm A}/S_{\rm C}>2500$. This improves on
the limits observed by \citet{win00a} by a factor of 8. The bright
images have different flux densities and different flux density ratios
in our six epochs, confirming the fluctuations seen by \citet{win03a}.
The larger fractional variation in image B relative to image A is
suggestive of scintillation. The possibility that the central image
also undergoes scintillation is a source of systematic uncertainty in
the interpretation of the upper limit on $S_{\rm C}$.

Lens galaxies at $z\lesssim0.5$ have isothermal profiles on scales of
a few kiloparsecs \citep{rus03a,koo06a}, and we find that the lens
galaxy in J1838 maintains this steep profile to within a few tens of
parsecs of its center. The lenses J1632--0033, B0739+366 and B1030+074
also have central density profiles close to isothermal
\citep{win03a,rus01a}. These four lens galaxies resemble the nearby
cusp galaxies in the photometric observations of
\citet{lau95a}. Galaxy simulations which track both dark matter and
baryons predict isothermal profiles at galactic radii $\gtrsim1$~kpc
(see Section~\ref{intro}), and central image constraints from these
galaxies show these profiles continuing to smaller radii.  A black
hole with a mass estimated from the $M-\sigma$ relation weakens the
constraints somewhat, allowing the central profile to be a little more
shallow.

As we add more galaxies to the ELVIS sample, we will have a larger
sample of gravitational lenses with either central images or very good
upper limits on the central image flux density.  Central images are
very sensitive to the central matter distributions that are just
shallower than the isothermal profiles of many local galaxies and of
distant galaxies at larger radii.  With more well-constrained central
matter profiles, we can determine if most early type galaxies at
$z\sim0.5$ have cuspy central profiles, or if some have shallow cores.
As well as the central image from the galaxy core, we may detect
central image pairs and directly measure black hole masses in distant
galaxies.

\acknowledgments

We thank David Rusin and Chuck Keeton for their participation in
ELVIS, and for their role in developing the lens modeling methodology
used in this paper. We also thank them, Shep Doeleman, and Scott Gaudi
for helpful consultations. We thank the referee for useful suggestions
that considerably improved this paper. Support for this work was provided 
by the National Science Foundation through grant AST 00-71181. E.B.\
acknowledges the support of an NRAO pre-doctoral fellowship.
The National Radio Astronomy Observatory is a facility of the 
National Science Foundation operated under cooperative agreement by
Associated Universities, Inc.

Facilities: NRAO (VLBA).

\bibliographystyle{apj}
\bibliography{apj-jour,elvis1_ap}

\end{document}